\documentstyle[aps,prc,psfig]{revtex}
\newcommand{\thalf}{\mbox{$\frac{3}{2}$}}
\newcommand{\ohalf}{\mbox{$\frac{1}{2}$}}
\newcommand{\coupp}{\mbox{\Large $\frac{f_{\pi N \Delta}}{m_{\pi}}$}}
\begin{document}
\title{Damping mechanisms of the $\Delta$ resonance in nuclei}
\author{B. K\"orfgen, P. Oltmanns\thanks{new address:
Institut f\"{u}r Niedertemperatur-Plasmaphysik, Robert-Blum-Str. 8-10,
17489 Greifswald, Germany}, and
F. Osterfeld}
\address{Institut f\"{u}r Kernphysik,
Forschungszentrum J\"{u}lich GmbH, D-52425 J\"{u}lich, Germany}
\author{T. Udagawa}
\address{Department of Physics, University of Texas, Austin, 
Texas 78712}
\maketitle 
\begin{abstract}
The damping mechanisms of the $\Delta$(1232) resonance in nuclei 
are  studied by analyzing 
the quasi-free decay reactions 
$^{12}$C($\pi^+$,$\pi^+$p)$^{11}$B and
$^{12}$C($^{3}$He,t$\pi^+$p)$^{11}$B and the
2p emission reactions
$^{12}$C($\pi^+$,pp)$^{10}$B and 
$^{12}$C($^{3}$He,tpp)$^{10}$B.
The coincidence cross sections
are calculated within the framework of the
isobar-hole model. It is found that the
2p emission process 
induced by the decay of the $\Delta$ resonance in the nucleus
can be consistently described  by a $\pi+\rho+g'$ model for the
$\Delta$+N$\rightarrow$ N+N decay  interaction.
\end{abstract}


\section{Introduction} 
Recently, new information on the $\Delta(1232)$-propagation in
the nucleus has been obtained from a
coherent  pion decay experiment~\cite{Hennino92,Hennino93}
and from a
$(\vec{p},\vec{n})$  spin-flip  transfer 
experiment~\cite{Prout94,Prout96}. In the  first experiment the
$^{12}$C($^{3}$He,t$\pi^+$)$^{12}$C(g.s.)
reaction  was used to measure
the isovector spin-longitudinal ($\vec{S}\cdot\vec{q}\;\vec{T}$)
response function in the $\Delta$ resonance region.
In the second experiment the spin observables of the
$(\vec{p},\vec{n})$ reaction
were used to decompose the charge exchange cross section
in the $\Delta$ resonance region into its
spin-longitudinal (LO) and spin-transverse (TR)  components.
Similar to the $\pi$-nucleus total cross section 
data~\cite{Binon70,Carroll76,Ericson88,Koerfgen94}
the LO cross sections of both reactions
show a substantial
downward energy shift of the $\Delta$ resonance in nuclei,
as compared to the proton target.
From a consistent $\Delta$-hole model
analysis of  pion and photon scattering, and charge exchange
reactions~\cite{Koerfgen94,Udagawa90,Delorme91,Udagawa94}
it is found that a large part of the observed shift
is due  to a nuclear medium effect on the LO response function.
The medium effect is caused by the strongly attractive,
energy dependent $\Delta$-particle - nucleon-hole residual interaction
$V_{\Delta N,\Delta N}$.
In Refs.~\cite{Koerfgen94,Udagawa90,Delorme91,Udagawa94,Oltmanns93}
it was shown that $V_{\Delta N,\Delta N}$
can be well described by the $\pi+\rho+g'$ model
(\cite{Osterfeld92} and references therein).
The strong attraction of the $\pi$-exchange
in the LO channel produces a collective pion mode at
excitation energies of $\sim 250$ MeV in the laboratory  frame.
The collectivity shifts the LO response function
down in energy by 60 MeV relative to the spin-transverse
($\vec{S}\times\vec{q}\;\vec{T}$) response function.
Other, smaller effects come from $\Delta$ conversion processes,
such as $\Delta N \rightarrow N+N$~\cite{Udagawa90,Delorme91},
and from projectile excitation~\cite{Oset89,Cordoba92}.
 
In the present paper we apply the
$\Delta$-hole model of Refs.~\cite{Koerfgen94,Udagawa94}
to the calculation of the damping of the collective pion mode 
in the nucleus. The major decay channels
are the coherent pion decay, the quasi-free decay, and
the 2p emission. While various calculations for
the coherent pion decay were published already in 
Refs.~\cite{Koerfgen94,Udagawa94,Oltmanns93,Cordoba93,Cordoba95}
we give here the results for the quasi-free ($\pi^+$p) decay
and the 2p emission. Since the coupling interaction for the
quasi-free decay is known we can use this process
to study the distortion effects on the
outgoing pion and proton wave functions.
For the 2p emission process we assume
a $\pi+\rho+g'$ interaction. We show
that the 2p emission in the $\Delta$ resonance energy region is
dominated by the zero-range Landau-Migdal interaction, 
the strength of which
can be exctracted from the data.
Both the pion induced reactions and the charge
exchange reactions are well reproduced by calculations
with a Landau-Migdal parameter in the range of 
$g'_{N\Delta}\approx 0.25 - 0.35$.

\section{Theory} 
In this section we describe the $\Delta$-hole model used
in the analysis of the experimental data. The formalism
and the methods of calculation were presented already in recent 
papers~\cite{Koerfgen94,Udagawa94}. In the present
paper we discuss only those formulas which are connected
with the quasi-free decay and the 2p emission.
\subsection{The excitation processes}
We start our formulation by writing down the
fivefold differential cross section
for the charge exchange reaction
$A+a\rightarrow (B+\Delta)+b\rightarrow C+c+d+b$
in the LAB frame ($E_A = m_A$, $\vec{p}_A =0$)
\begin{eqnarray}
\frac{d^5\sigma}{dE_b d\Omega_b dE_cd\Omega_c d\Omega_d}
& = &
\frac{1}{(2\pi)^8}
\frac{1}{\mid \vec{v}_{rel}\mid}
\frac{m_a}{E_a}\,
\frac{m_b}{E_b}\,
\frac{m_c}{E_c}\,\frac{m_d}{E_d}\,\frac{m_C}{E_C}\,
\frac{p_b\, E_b\; p_{c}\, E_c\; p_{d}\, E_d\; E_C}
{\mid(m_A+\omega_{c.e.}-E_c)-\frac{E_{d}}{p_{d}}
(p_a \cos{\theta_{ad}}
-p_b \cos{\theta_{bd}}-p_c\cos{\theta_{cd}})\mid}
\nonumber \\ & & \times
\overline{\sum}\mid T_{fi}\mid^2 \;\; .
\label{eq1}
\end{eqnarray}
Here $A$ ($B+\Delta$) and $a$ ($b$) denote the target
(excited intermediate nucleus) and projectile (ejectile), 
respectively. The $B+\Delta$ system
de-excites to the residual nucleus $\mid\varphi_{C}\rangle$ by
emission of the particles $c$ and $d$ which carry four-momenta
($E_{c},\vec{p}_{c})$ and ($E_{d},\vec{p}_{d})$, respectively.
$m_i$ stands for the mass of particle $i\;(i=A,C,a,b,c,d$) and
($\omega_{c.e.},\vec{q}_{c.e.}$) denotes the four-momentum transfer
in the excitation process.
In case that one of the outgoing particles is a boson (e.g. in the
quasi-free decay) the according normalisation factor  M/E has to
be replaced by 1/2E. The full four body kinematics in the final
reaction channel is included.
 
The transition amplitude $T_{fi}$
for the decay process is defined as
\begin{equation}
T_{fi}=\langle\varphi_C, \left [
\phi_c(\vec{p}_{c})\;\phi_d(\vec{p}_{d}) \right ]_{({\cal A})}
\mid V_{cd,\Delta}
\mid \psi\rangle
\label{eq2}
\end{equation}
where $V_{cd,\Delta}$ denotes the
$\Delta$ decay interaction that will be specified later.
$\phi_c(\vec{p}_{c})$ and $\phi_d(\vec{p}_{d})$
are the distorted wave functions of the 
outgoing particles $c$ and $d$, respectively,
and $\varphi_C$ is the wave function of the residual nucleus.
The index $({\cal A})$ indicates that in case of the 2p emission
the wave functions of the two outgoing identical fermions have to be 
antisymmetrized. The  wave function $\mid\psi\rangle$ describes
the intermediate $B+\Delta$ system and is defined by~\cite{Udagawa94}
\begin{equation}
\mid\psi\rangle  =  G\mid\rho\rangle   =
\frac{1} {\omega +i\Gamma_{\Delta} /2-H_B-T_{\Delta}-U_{\Delta}-
V_{\Delta N, \Delta N}}\mid \rho\rangle
\label{eq3}
\end{equation}
where $\mid \rho\rangle$ is the doorway state
excited initially by the reaction.
The Green's function $G$  describes
the propagation of the ($B+\Delta$) system and
is approximated by that of the isobar-hole 
model~\cite{Hirata77,Hirata79,Horikawa80,Oset82}.
$\Gamma_{\Delta}(\omega)$ 
is the energy dependent free decay width of the $\Delta$,
$H_{B}$ is the Hamiltonian of the hole nucleus $B$,
$T_{\Delta}$ and $U_{\Delta}$ are the kinetic energy operator
and the $\Delta$-nucleus  one-body potential, respectively, and
$V_{\Delta N,\Delta N}$ is the $\Delta$-hole residual interaction.
For the calculation of $|\psi\rangle $ we use the same input
parameters as used in Refs.~\cite{Koerfgen94,Udagawa94}.
The Pauli blocking effects are assumed to be included in the
average $\Delta$-nucleus one-body potential, which we fixed
by re-analysing the relevant scattering data~\cite{Binon70,Carroll76}.
We refer the reader for more details to 
Refs.~\cite{Koerfgen94,Udagawa94}.

For charge exchange reactions the doorway state $\mid \rho\rangle$
has the explicit form~\cite{Udagawa94}
\begin{equation}
\mid\rho_{c.e.}\rangle=(\chi_b^{(-)}\varphi_b\mid
 t_{NN,N\Delta}
 \mid\chi_a^{(+)}\varphi_a \varphi_A\rangle   \;\;
\label{eq4}
\end{equation}
where $\chi_a^{(+)}$ and $\chi_b^{(-)\ast}$ denote the projectile
and ejectile distorted wave functions, 
respectively, and $\varphi_{a}$ and $\varphi_{b}$ are  the
corresponding intrinsic wave functions of the projectile
$a$ and ejectile $b$;
$\mid\varphi_A\rangle$ describes the target ground state
wave function. The effective
$NN\rightarrow N\Delta$ transition operator for the charge
exchange process is denoted by $t_{NN,N\Delta}$.
The round bra (  $\; \mid$ on the right side of eq.~(\ref{eq4})
denotes the integration with
respect to the projectile coordinates only.

For pion induced reactions the coincidence cross section in the LAB 
frame is threefold differential
\begin{equation}
\frac{d^3\sigma}{dE_cd\Omega_c d\Omega_d}
= \frac{1}{(2\pi)^5}\,
\frac{1}{\mid \vec{v}_{rel}\mid}\,
\frac{1}{2\omega_{\pi}}\,
\frac{m_c}{E_c}\,\frac{m_d}{E_d}\,\frac{m_C}{E_C}\,
\frac{p_{c}E_c p_{d}E_d E_C}
{\mid(m_A+\omega_{\pi}-E_c)-\frac{E_{d}}{p_{d}}
(p_a \cos{\theta_{ad}}
-p_c\cos{\theta_{cd}})\mid}
\overline{\sum}\mid T_{fi}\mid^2
\label{eq5}
\end{equation}
where ($\omega_{\pi},\vec{q}_\pi$) is the four-momentum of the 
incident pion. Now  the doorway state has the form~\cite{Koerfgen94}
\begin{equation}
\mid\rho_{\pi}\rangle =
\frac{f_{\pi N\Delta}}{m_\pi}
\; \biggl ( \vec{q}_\pi \cdot \vec{S}^\dagger
\biggr ) \;\; T^\dagger_{\nu} \;\; e^{i\vec{q}_\pi \cdot \vec{r}}
\mid\varphi_A\rangle
\label{eq6}
\end{equation}
where $\vec{S}^\dagger$ and $\vec{T}^\dagger$ are the
spin and isospin transition operators, respectively, that
convert a nucleon into a $\Delta$(1232) isobar.
The coupling constant $f_{\pi N\Delta}$ is fixed from pion-nucleon
scattering data and has the value
$f^2_{\pi N\Delta}/4\pi = 0.324$. The index $\nu=\pm 1$
distinguishes between $\pi^{\pm}$ scattering.

\begin{figure}
\centerline{\psfig{figure=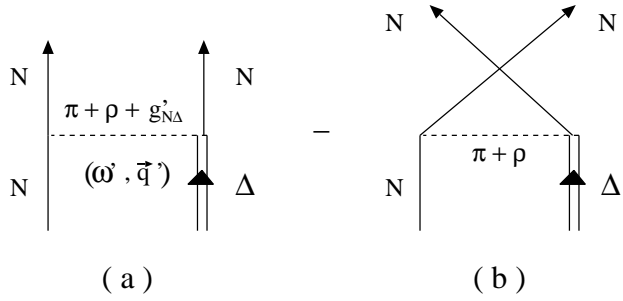,height=4.0cm,angle=-90}}
\vskip0.4cm
\caption{Schematic representation of the 2p emission 
induced by the decay of the 
$\Delta$: (a) direct graph, (b) exchange graph.} 
\end{figure}

\subsection{The decay interactions}
\label{decay} 
In case of the $^{12}$C($\pi^+$,$\pi^+$p)$^{11}$B and
$^{12}$C($^{3}$He,t$\pi^+$p)$^{11}$B reactions
the decay interaction $V_{cd,\Delta}$ of eq.~(\ref{eq2}) is 
represented by
\begin{equation}
V_{p\pi ,\Delta} \;\; = \;\; \frac{f_{\pi N\Delta}}{m_{\pi}}\,
\vec{S}\cdot\vec{q}^{\,\prime}_{\pi}\,
T_{\mu} .
\label{eq7}
\end{equation}
Note that this interaction has no free
parameter and is known from elastic pion scattering in the
$\Delta$ resonance region.
The explicit formulas for the quasi-free decay of the $\Delta$ are
given in appendix~\ref{app1}.

In case of the $^{12}$C($\pi^+$,pp)$^{10}$B
and $^{12}$C($^{3}$He,tpp)$^{10}$B reactions
the interaction for the process $\Delta+N\rightarrow N+N$
is described by a $\pi +\rho +g'$ model~\cite{Osterfeld92}:
\begin{equation}
V_{pp,\Delta}(\omega^{\prime} , \vec{q}^{\,\prime})\;\; =\;\; V_{\pi} 
(\omega^{\prime} ,\vec{q}^{\,\prime})
\; +\; V_{\rho}(\omega^{\prime} ,\vec{q}^{\,\prime}) \; + 
\; V_{\delta} \;\; ,
\label{eq8}
\end{equation}
with
\begin{equation}
V_{\delta}\;\; =\;\;
\hbar c\; \frac{f_{\pi NN}f_{\pi N\Delta}}{m^{2}_{\pi}} \; 
g'_{N\Delta} \;
\left ( \vec{\sigma}_{2} \cdot \vec{S}_{1} \right ) \, 
\left ( \vec{\tau}_{2}\cdot \vec{T}_{1} \right ) \;\; .
\end{equation}
In eq.~(\ref{eq8}),
$\omega^{\prime}$ and $\vec{q}^{\,\prime}$ are the energy and
three-momentum transfer involved in the interaction 
$\Delta+N\rightarrow N+N$, respectively.
The interaction $V_{\delta}$ is the so-called 
Landau-Migdal term. 
It describes the short range correlations for
$\Delta +N\rightarrow N+N$ transitions. The special value 
for the Landau-Migdal parameter $g'_{N \Delta} = 1/3$
(in units of $\hbar c f_{\pi NN}f_{\pi N\Delta}/m_{\pi}^2 \approx$
800 MeV fm$^3$) is known as the `minimal $g'_{N\Delta}$'
because it cancels out the attractive short range part of the
$\pi$-echange potential~\cite{Osterfeld92}.

The $\pi$- and $\rho$-exchange potentials $V_{\pi}$ and $V_{\rho}$ are 
defined consistently with the potentials for the residual interaction:
\begin{equation}
V_{\pi}(\omega^{\prime} ,\vec{q}^{\,\prime}) = \hbar c \;
\frac{f_{\pi NN}f_{\pi N\Delta}}{m^2_{\pi}}\,
\left ( \frac{\Lambda_{\pi}^2-m_{\pi}^2}{\Lambda_{\pi}^2-t^{\prime}} 
\right ) ^2 \; \frac{1}{t^{\prime}-m_{\pi}^2+i\varepsilon} \; 
(\vec{\sigma}_{2}\cdot \vec{q}^{\,\prime}) 
(\vec{S}_{1}\cdot \vec{q}^{\,\prime}) \; 
\left ( \vec{\tau}_{2}\cdot \vec{T}_{1} \right )
\label{eq9}
\end{equation}
\begin{equation} 
V_{\rho}(\omega^{\prime} ,\vec{q}^{\,\prime})  =  \hbar c\;
\frac{f_{\rho NN}f_{\rho N\Delta}}{m^2_{\rho}}\,
\left ( \frac{\Lambda_{\rho}^2-m_{\rho}^2}{\Lambda_{\rho}^2-t^{\prime}} 
\right ) ^2 \; \frac{1}{t^{\prime}-m_{\rho}^2+i\varepsilon} \; 
(\vec{\sigma}_{2} \times \vec{q}^{\,\prime}) 
\cdot (\vec{S}_{1}\times \vec{q}^{\,\prime})\; 
\left ( \vec{\tau}_{2}\cdot \vec{T}_{1} \right )
\; .
\label{eq10}
\end{equation}
In the eqs.~(\ref{eq9}) and (\ref{eq10}), 
$t^{\prime}=\omega^{\prime 2} -\vec{q}^{\,\prime 2}$
is the four-momentum transfer in the decay process, 
$m_{\pi}$ and
$\Lambda_{\pi}$ ($m_\rho$, $\Lambda_\rho$) are the mass and
cutoff mass of the $\pi$ ($\rho$), respectively.
The various parameters are fixed as follows:
$f_{\pi NN}f_{\pi N\Delta}/4\pi =0.162$, 
$f_{\rho NN}f_{\rho  N\Delta}/4\pi =8.32$,
$m_{\pi}=0.14$ GeV, $m_{\rho}=0.77$ GeV,
$\Lambda_{\pi}=1.20$ GeV, and $\Lambda_{\rho}=2$ GeV.

As a consequence of the Pauli principle the wave functions of the 
two outgoing protons have to be antisymmetrized. This leads to two 
contributions to the 2p emission
process, namely the direct and the exchange term
(see Fig.~1). The antisymmetrization has to be carried out only for 
the finite-range $\pi$- and $\rho$-exchange potentials. 
The Landau-Migdal term is a zero-range 
interaction and  has not to be antisymmetrized.
This treatment of the 
$\Delta+N\rightarrow N+N$ interaction is in line with 
microscopic nuclear structure calculations~\cite{Meyer81,Towner87}
and microscopic G-Matrix calculations~\cite{Krewald88}.
Therefore the Landau-Migdal
parameter $g'_{N\Delta}$ extracted from the 2p emission  
reactions can be directly compared with the values found 
in these calculations~\cite{Meyer81,Towner87,Krewald88}. 
The explicit formulas for the 2p emission process induced by the 
decay of the $\Delta$ in the nucleus are 
derived in appendix~\ref{app2}.

\begin{figure}
\centerline{\psfig{figure=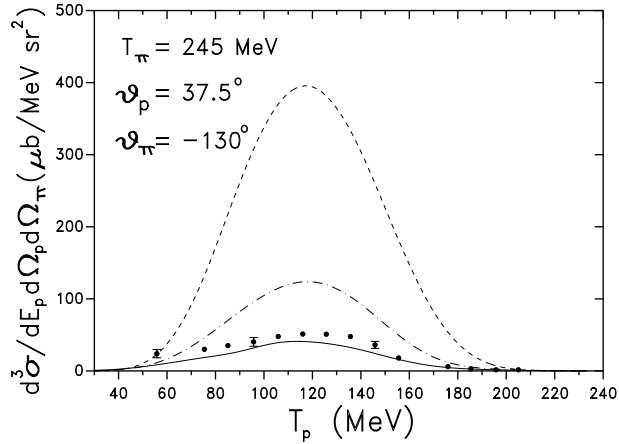,height=6.0cm,angle=90}}
\vskip 0.3cm
\caption{Triple differential cross section of the
$^{12}$C($\pi^+$,p$\pi^+$)$^{11}$B reaction at given pion and
proton angle plotted versus
the kinetic energy of the decay proton.
The data have been taken from 
Ref.~\protect\cite{Piasetzky82}.
The dashed curve shows the result without
inclusion of the residual interaction
$V_{\Delta N,\Delta N}$
while the dot-dashed curve shows the result with
the inclusion of
$V_{\Delta N,\Delta N}$.
The solid curve shows the result with additional inclusion of
the distortion effect on the decay pion and proton wave functions.}
\end{figure}

\section{Results and Discussion} 
With the formalism described in Sec.~II we have calculated
cross sections  for the quasi-free decay reactions
$^{12}$C($\pi^+$,$\pi^+$p)$^{11}$B  and 
$^{12}$C($^3$He,t$\pi^+$p)$^{11}$B
and the 2p emission reactions
$^{12}$C($\pi^+$,$\pi^+$pp)$^{10}$B  and 
$^{12}$C($^3$He,t$\pi^+$pp)$^{10}$B.
While the pion induced reactions are truely exclusive with
the decay particles measured in coincidence at given energies
and angles  the decay cross sections of the
charge exchange reaction 
have been integrated over a certain
kinematical range, as determined by the geometry of the
DIOGENE detector~\cite{Hennino92,Hennino93}.
Therefore the integrated cross sections 
for ($^3$He,t) induced processes are in a way less exclusive
than the pion induced reactions and show only the gross 
features of the process. 

Furthermore, our calculations  treat
the distorsion effects on the incoming
and outgoing particles in the adequate frameworks.
In the pion induced reaction the distorsion of the incoming
pion is treated within the isobar hole model while the distorsion
of the decay pion and protons  is described by optical model
wave functions~\cite{Becchetti69,Eisenstein74}. 
In the ($^3$He,t) charge exchange reaction
the projectile and ejectile and the decay particles are
described by optical model wave functions, whereas
the $\Delta$ propagation through the nucleus is again treated within
the isobar-hole model.

\subsection{The reaction $^{12}$C($\pi^+$,p$\pi^+$)$^{11}$B}
In Fig.~2 we compare the results of our calculations
for the $^{12}$C($\pi^+$,p$\pi^+$)$^{11}$B reaction
to the data of Ref.~\cite{Piasetzky82}
at the  pion kinetic energy of $T_{\pi}=245$ MeV.
In the experiment the angles of the outgoing proton and pion were
fixed at  $\theta_p=37.5^{\circ}$ and
$\theta_{\pi}=-130^{\circ}$, respectively, in coplanar geometry.
The threefold differential cross section is plotted versus
the kinetic energy of the outgoing proton.
Three different calculations are compared to the data.
The dashed curve shows the result of our calculations
with $V_{\Delta N,\Delta N}=0$ while the dot-dashed
curve shows the result with $V_{\Delta N,\Delta N}\neq 0$.
By comparison of both curves one recognizes
that the inclusion of the residual interaction
reduces the quasi-free decay cross section
by  a factor of $\sim 4$.
This reduction is due to the absorption taking place in the
multiple scattering of the pion. 
A similar reduction factor is also observed in the
total pion-nucleus cross section~\cite{Koerfgen94}.
The solid curve shows the result with additional
inclusion of the distortion effect on the outgoing
proton and pion wave functions.
The distortion effect leads to a further reduction
of the cross section by a factor of $\sim 3$.
We describe the  relative motion of the decay particles with
respect to the residual nucleus by optical model wave functions.
This is a consistent method within the framework of direct nuclear
reaction theory and has been used in the analysis of other
reactions, like A(e,e' pp).
For the calculation of the proton and pion wave functions
we used the optical potential parameters, as derived from
elastic proton-nucleus~\cite{Becchetti69} 
and elastic pion-nucleus scattering~\cite{Eisenstein74}.
Using these optical model wave functions we
overestimate the absorption and thus obtain a lower limit for the
$^{12}$C($\pi^+$,p$\pi^+$)$^{11}$B cross section.
A comparison of the solid curve with the data shows that we 
underestimate the data by $\sim 10 - 20$\%~. This is in 
agreement with our expectation and assures us that our treatment of 
the distortion effects is
reasonable. In the following calculations we will use the same model
for the description of the distortion effects on the outgoing particles.

\begin{figure}
\centerline{\psfig{figure=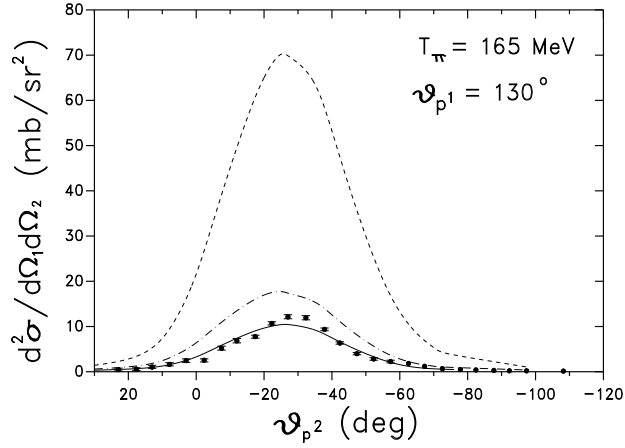,height=6.0cm,angle=90}}
\vskip 0.3cm
\caption{Coincidence spectra for the $^{12}$C($^3$He,t) reaction
at $T_{He} = 2$ GeV and triton scattering angle $\theta_t =0^\circ$.
(a) The $^{12}$C($^3$He,tp$\pi^+$) data~\protect\cite{Hennino93}
in comparison with the result of our calculation.
(b) The $^{12}$C($^3$He,tpp) reaction data~\protect\cite{Hennino93}.
The solid and dashed curve show the results with
and without inclusion of the residual interaction, respectively.}
\end{figure}
\begin{figure}
\centerline{\psfig{figure=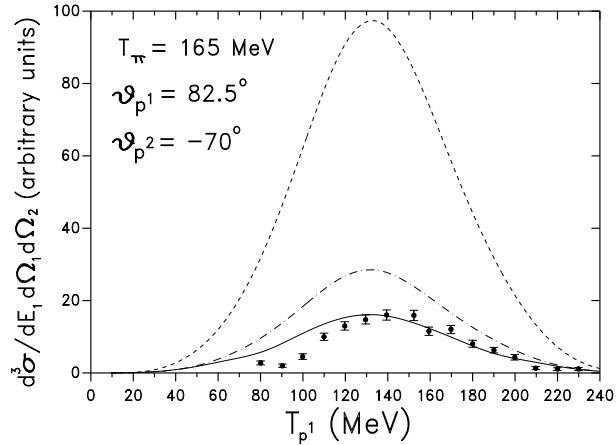,height=6.0cm,angle=90}}
\vskip 0.3cm
\caption{Double differential cross section of the
$^{12}$C($\pi^+$,pp)$^{10}$B reaction. The angle of proton 1 is fixed 
at $\theta_{p^1}=130^{\circ}$. The cross section is plotted
versus the angle of proton 2.
The data have been taken from Ref.~\protect\cite{Altman83,Altman86}.
The dashed curve shows the result with $V_{\Delta N,\Delta N}=0$
while the dot-dashed curve shows the result with
$V_{\Delta N,\Delta N}\neq 0$.
The solid curve shows the result with additional inclusion of
the distortion of the decay protons.}
\end{figure}

\subsection{The reaction $^{12}$C($\pi^+$,pp)$^{10}$B} 
In Fig.~3 we show  the results for the 
$^{12}$C($\pi^+$,pp)$^{10}$B cross section at $T_{\pi}$ = 165 MeV.
The experiment was performed in coplanar geometry with
the angle of proton 1 fixed at $\theta_{p^1}=130^{\circ}$.
The threefold differential cross section is plotted versus
the angle of proton 2.
The dashed curve shows the result of our calculation
with $V_{\Delta N,\Delta N}=0$ while the dashed-dotted curve shows
the result with $V_{\Delta N,\Delta N}\neq 0$.
The inclusion of the multiple scattering of the pion in the medium
results again in a reduction of the cross section by  a factor 
of $\sim 4$.
The solid curve shows the result with additional
inclusion of the distortion effect on the outgoing
proton wave functions. This leads to a further
reduction of the cross section by a factor of $\sim 2$.
The calculations shown here were performed  with a minimal 
Landau-Migdal parameter $g'_{N\Delta} = 1/3$ in the 2p emission
matrix element,
which gives a very good agreement with the data~\cite{Altman83}.
Neglection of the $\pi$- and $\rho$-exchange potentials in the
decay interaction leads to almost the same results.
Thus the Landau-Migdal
term is the most important ingredient to the $\Delta+N\rightarrow N+N$
interaction. This is a consequence of the large momentum and energy
transfer involved in the $\Delta+N\rightarrow N+N$
decay process leading to a very short ranged interaction.  
We remark that the determination of the $g'_{N\Delta}$ parameter in
the ($\pi^+$,pp)-reaction is rather direct since this reaction
is dominated by the intermediate $\Delta^{++}$ excitation. On the other 
hand, in case of the $^{12}$C($\pi^+$,pn)$^{10}$C 
reaction~\cite{Altman86} or photon induced 2p 
emission reactions~\cite{Machenil93,Ryckebusch94}
many competing processes are possible and the
Landau-Migdal parameter $g'_{N\Delta}$ can only be extracted
in an indirect way.

In Fig.~4 we study the same reaction as in Fig.~3; here
the two protons were detected at
$\theta_{p^1}=82.5^{\circ}$ and $\theta_{p^2}=-70^{\circ}$,
and the cross section is plotted
as function of the kinetic energy of proton 1. 
The curves shown have the same meaning as in Fig.~3.
Note that the data are given in arbitrary units. 
Thus the data are normalized to the solid curve. 
The inclusion of the residual interaction and the inclusion
of the distortion effects reduces the magnitude again by a factor 
of $\sim 4$ and $\sim 2$, respectively.
The dependence of the cross section on the kinetic energy of the 
proton 1 is reproduced well. The 
slight deviation of the calculated cross section from the data 
at low kinetic energies could indicate 
that the treatment of the distortion on the outgoing nucleons 
should be improved for these energies.   

\subsection{The reaction  $^{12}$C($^3$He,tp$\pi^+$)}
In Fig.~5~(a) we compare our results for the
 $^{12}$C($^3$He,tp$\pi^+$) reaction with
the data of Hennino {\it et al.}\/~\cite{Hennino93}.
The data are integrated over the kinematically allowed
pion and proton energies as well as pion and proton angles.
The solid curve shows our calculation with inclusion of the residual
interaction $V_{\Delta N,\Delta N}$ and with inclusion of
the distortion effects on the outgoing particles.
We see  no effect of the
residual interaction in this energy and angle integrated
cross section, i. e. the cross sections with
and without inclusion of $V_{\Delta N,\Delta N}$ are the same.
This is an indication that the p$\pi^+$ events in this reaction 
come only from the nuclear surface. 

\begin{figure}
\centerline{\psfig{figure=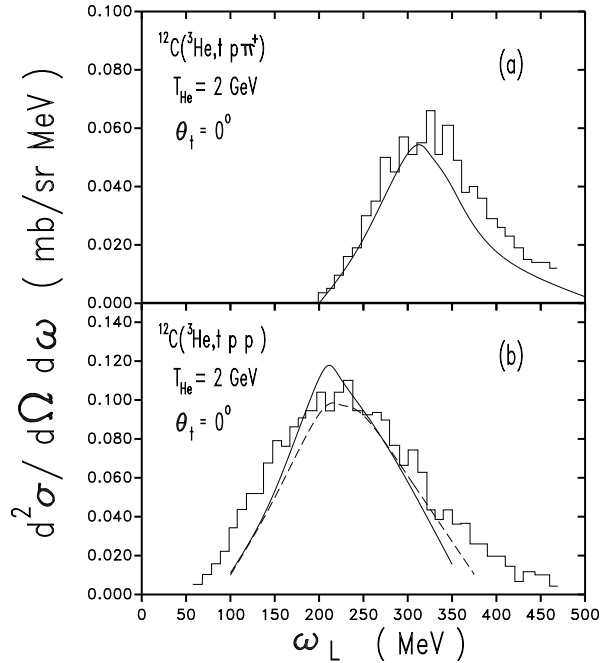,height=9.0cm}}
\vskip 0.3cm
\caption{Threefold differential cross section of the
$^{12}$C($\pi^+$,pp)$^{10}$B reaction. The two protons were detected
at $\theta_{p^1}=82.5^{\circ}$ and $\theta_{p^2}=-70^{\circ}$, 
respectively. The cross section is plotted
as function of the kinetic energy of proton 1. 
The data have been taken from Ref.~\protect\cite{Altman83,Altman86}.
The curves shown have the same meaning as in 
Fig.~3. Note that the data are given in arbitrary units and 
are thus normalized to the solid curve.}
\end{figure}

\subsection{The reaction $^{12}$C($^3$He,tpp)} 
In Fig.~5~(b) we compare our microscopically calculated
$^{12}$C($^3$He,tpp) coincidence  spectra
with the data of Hennino {\it et al.}\/~\cite{Hennino93}.
The data are integrated over the phase space of both outgoing protons.
Two calculations are compared to the data: The solid curve shows
the result with $V_{\Delta N,\Delta N}\neq 0$
while the dashed curve shows the result with 
$V_{\Delta N,\Delta N}=0$.
In both cases the distortion effects are taken into account.
In order to reproduce the magnitude of the experimental data
we used a Landau-Migdal parameter of $g'_{N\Delta} = 0.28$.
This value is consistent with
that found in the ($\pi^+$,pp)-reactions of Figs.~3 and 4.
Both values for $g'_{N\Delta}$ lie in the range from 0.25 -- 0.35
and are thus in agreement with values of the Landau-Migdal
parameter found in microscopic G-Matrix calculations~\cite{Krewald88}.

\section{Summary} 
In conclusion, we have presented microscopic calculations
for the quasi-free decay of the $\Delta$ resonance 
and the 2p emission in nuclei
induced by pion
absorption and by charge exchange reactions. 
These calculations are performed within the framework of
the $\Delta$-hole model and are
consistent with our former calculations of inclusive
and exclusive reactions on nuclei.
Since the coupling interaction for the
quasi-free decay is known we use this process
to study the distortion effects on the 
outgoing pion and proton wave functions.
In the  2p emission reactions we describe the
$\Delta$+N$\rightarrow$ N+N decay  interaction
by a $\pi +\rho +g'$ model, which is
consistent with our description of the residual
interaction. We find that the $\Delta+N\rightarrow N+N$ 
decay interaction 
is dominated by the zero-range Landau-Migdal term.
The data for both the ($\pi^+$,pp) reaction
and the ($^3$He,tpp) reaction
are well reproduced by calculations
with a Landau-Migdal parameter in the range of 
$g'_{N\Delta}\approx 0.25 - 0.35$.
 
This work is supported in part by the U.S. Department of Energy
under Contract DE-FG05-84-ER40145.

\appendix
\section{Explicit formulas for the quasi-free decay}
\label{app1}
In this appendix we derive the explicit formulas for the transition 
amplitude of the quasi-free decay. 
Inserting the decay interaction $V_{p\pi ,\Delta}$
of eq.~(\ref{eq7}) into eq.~(\ref{eq2}) 
one can write the transition amplitude in the following way:
\begin{equation}
T_{q.f.} = \langle J^{'}_{h}M^{'}_{h}
; \vec{q}^{\,\prime}_{p} \frac{1}{2}\, m^{\prime}_s ; 
\vec{q}^{\,\prime}_{\pi} \mid
\frac{f_{\pi N\Delta}}{m_{\pi}}\,
\vec{S}\cdot\vec{q}^{\,\prime}_{\pi}\,
T_{\mu}\mid \psi \rangle
\label{AGl1}
\;\; .
\end{equation}
Here $J^{'}_{h}$ and $M^{'}_{h}$ represent the total angular
momentum and the magnetic quantum number of the final (A-1)-nucleus;
$\vec{q}^{\,\prime}_{p}$ and $m^{\prime}_s$ are the three-momentum 
and the spin projection
of the outgoing proton; $\vec{q}^{\,\prime}_{\pi}$ is the 
three-momentum of the outgoing pion.
We expand the wave function $\mid \psi \rangle$ of the 
$\Delta$-hole state
in terms of the channel wave functions~\cite{Udagawa94}
\begin{equation}
\mid [Y_{\Delta}\Phi_{h}]_{j_t m_t} \rangle
 = \sum_{m_{\Delta}m_{h}}
(j_{\Delta}\, m_{\Delta}\; j_{h}\, m_{h}\, | \, j_t\, m_t )
\; \mid Y_{j_{\Delta}m_{\Delta}}\Phi_{j_{h}m_{h}}\rangle \; ,
\label{AGl2}
\end{equation}
where $Y_{j_{\Delta} m_{\Delta}}$ is the spin-angle wave function of
the $\Delta$ and $\Phi_{j_h m_h}$ is the hole wave function
of nucleus B. Thus we obtain~\cite{Udagawa94}
\begin{equation}
\label{AGl3}
\mid \psi \rangle =  \sum_{j_{t}m_t}\,
 \sum_{\Delta h}^{N_{c}} \; \psi^{(j_t m_t)}_{\Delta h}(r)\, 
\frac{1}{r} \,\mid [Y_{\Delta}\Phi_{h}]_{j_t m_t}\rangle \; .
\end{equation}
In (\ref{AGl3}) $N_c$ denotes the total number of allowed 
$\Delta$-hole states.
The radial wave function is then given by the inversion of 
eq.~(\ref{AGl3}):
\begin{equation}
\label{AGl4}
\psi^{(j_t m_t)}_{\Delta h}(r) =  r \;
( [Y_{\Delta}\Phi_{h}]_{j_t m_t} \mid \psi \rangle \; .
\end{equation}
The wave functions of the decay nucleon and the outgoing pion are 
also expanded  in multipoles:
\begin{equation}
\mid \vec{q}^{\,\prime}_p \frac{1}{2}\, m^{\prime}_s\rangle
 =   4\pi \sum_{l_p m_{l_p}}\,\sum_{j_p m_p}
i^{l_p}\, \chi_{l_p}(q^{\prime}_p r)\, (l_p\, m_{l_p}\;
\ohalf\, m^{\prime}_s\mid j_p\, m_p )\,
Y^{\ast}_{l_p m_{l_p}} (\hat{q}^{\prime}_p ) \; \left [
Y_{l_p}(\hat{r})\otimes
\chi_{\frac{1}{2}} \right ]_{j_p m_p} \;\; ,
\label{AGl6}
\end{equation}
\begin{equation}
\mid \vec{q}^{\,\prime}_{\pi}\rangle
 =  4\pi \sum_{l_{\pi} m_{l_{\pi}}}
i^{l_{\pi}}\, \chi_{l_{\pi}}(q^{\prime}_{\pi} r)\, 
Y^{\ast}_{l_{\pi} m_{l_{\pi}}} (\hat{q}^{\prime}_{\pi} ) \,
Y_{l_{\pi} m_{l_{\pi}}}(\hat{r}) \;\; ,
\label{AGl7}
\end{equation} 
where $\chi_{l_p}$ and $\chi_{l_\pi}$ denote the distorted radial
wave functions of the outgoing proton and pion, respectively.
Using the expansion of eq.~(\ref{AGl3}) 
for the wave function $\mid \psi\rangle$ and using the expansions
of eqs.~(\ref{AGl6}) and (\ref{AGl7})
for the outgoing nucleon and pion wave functions 
we can rewrite the transition amplitudes for
the quasi-free decay of the $\Delta$ as
\begin{eqnarray}
T_{q.f.}& = &
\sum_{j_t m_t}\,
\sum^{N_c}_{\Delta h}
\delta_{J^{'}_h , j_h} 
\delta_{M^{'}_h , m_h}
\sum_{l_p m_{l_p}}\,\sum_{j_p m_p}
\;\sum_{l_\pi j_\pi m_\pi}\,
2\sqrt{4\pi}^{\, 3}\;\coupp\; i^{l_\Delta + l_\pi - l_p}\;
\hat{\jmath}_\Delta\,\hat{l}_\Delta\, \hat{l}^{2}_\pi
\nonumber\\
&  & 
\;Y_{l_p m_{l_p}} (\hat{q}^{\,\prime}_p ) \;
(-1)^{j_\pi - m_\pi}\; Y_{j_\pi m_\pi}(\hat{q}^{\,\prime}_\pi )\; 
(j_\Delta\, m_\Delta\; j_h\, m_h \mid j_t\, m_t ) \; 
(j_\Delta \, m_\Delta\; j_\pi\, -m_\pi \mid j_p\, m_p )\;
\left \{ \begin{array}{ccc} l_p & \frac{1}{2} & j_p \\   \\
                            l_\Delta & \frac{3}{2} & j_\Delta \\   \\
                            l_\pi   &           1 & j_\pi
                            \end{array} \right \} 
\nonumber\\
& & 
(l_\Delta \, 0 \; l_\pi\, 0\mid l_p\, 0)\;
(l_\pi \, 0 \; 1 \, 0 \mid j_\pi\, 0 )\;
(l_p\, m_{l_p}\, \ohalf\, m^{\prime}_s\mid j_p\, m_p )\;
(1\, \mu \; \ohalf\,\tau_p\mid \thalf\, \tau_\Delta )\;
\int dr\; (q^{\prime}_\pi r)\;\chi_{l_p}(q^{\prime}_p r)\; 
\chi_{l_\pi}(q^{\prime}_\pi r)\;
\psi^{\; (j_t\, m_t)\,}_{(\Delta\, h)}(r)
\; .
\nonumber\\
\label{AGl5}
\end{eqnarray}
In eq.~(\ref{AGl5})  
the Clebsch-Gordan coefficient 
$(1\, \mu \; \ohalf\,\tau_p\mid \thalf\, \tau_\Delta )$
describes the isospin coupling coefficient of the decay process
and $\hat{x}=\sqrt{2x+1}$.

\section{Explicit formulas for the 2p emission} 
\label{app2}
Here we show the explicit formulas for the 2p emission processes.
Due to the antisymmetrization of the two outgoing protons the 
transition  amplitude for the 2p emission consists of the sum
of the direct and exchange transition amplitudes.
As a consequence of the arguments given in section~\ref{decay} we
have to antisymmetrize only the $\pi$- and $\rho$-exchange
interactions. The transition amplitude is given by (see Fig.~1):
\begin{eqnarray}
\label{BGl1}
T_{2N} & = & \langle J^{'}_{h}M^{'}_{h}
;\left [ \vec{q}^{\,\prime}_{p^1} \frac{1}{2}\, m^{\prime}_{s^1} ,
\vec{q}^{\,\prime}_{p^2} \frac{1}{2}\, m^{\prime}_{s^2}\right ] 
_{\cal A}
\mid V_{pp,\Delta} \mid \psi \rangle
\\
& = & \frac{1}{\sqrt{2}}\; \left ( \langle J^{'}_{h}M^{'}_{h}
; 1 ; 2 \mid V_{\pi} +V_{\rho} + V_{\delta}\mid \psi \rangle \; - \;
\langle J^{'}_{h}M^{'}_{h}
; 2 ; 1 \mid V_{\pi}+V_{\rho} \mid \psi \rangle \right )\;\; .
\label{BGl2}
\end{eqnarray}
In eqs.~(\ref{BGl1}) and (\ref{BGl2}) $J^{'}_{h}$ and $M^{'}_{h}$ 
represent the total angular
momentum and the magnetic quantum number of the final (A-2)-nucleus;
$\vec{q}^{\,\prime}_{p^1}$ and $m^{\prime}_{s^1}$
($\vec{q}^{\,\prime}_{p^2}$, $m^{\prime}_{s^2}$) 
are the three-momentum and the spin projection
of the outgoing proton 1 (2), respectively.
Using the expansion of eq.~(\ref{AGl3}) 
for the wave function $\mid \psi\rangle$ and the multipole expansion
of the nucleon wave functions of eq.~(\ref{AGl6}) we can write 
the matrix element for the direct decay graph in the following way:
\begin{eqnarray}
\langle J^{'}_{h}M^{'}_{h}
; 1 ; 2
\mid V_{pp,\Delta} \mid \psi \rangle
& = &
\sum_{j_t m_t}\,
\sum^{N_c}_{\Delta h}\,
\sum_{l_{p^1} m_{l_{p^1}}}\,\sum_{j_{p^1} m_{p^1}}\,
\sum_{l_{p^2} m_{l_{p^2}}}\,\sum_{j_{p^2} m_{p^2}}\,
\sum_{j^{\prime}_{h^2}\tau^{\prime}_{h^2}}\,
\sum_{J_p M_p}\,\sum_{J_1 J_2}\,\sum_{l_1 l_2}\;
2\,\sqrt{6}\; (4\pi ) 
\nonumber\\
& & \nonumber \\
& &
i^{l_1 + l_2 - l_{p^1}-l_{p^2}+l^{\prime}_{h^2}+l_\Delta}\;
(-1)^{J_1 + J_2 - j_t}\;
(-1)^{J_2 + M_2}\;
\hat{J}^{2}_1 \,\hat{J}^{2}_2\,\hat{J}_p\,
\hat{J}^{\prime}_h\, 
\hat{\jmath}^{\prime}_{h^2}\,\hat{\jmath}_{p^1}\,
\hat{\jmath}_{p^2}\,
\hat{\jmath}_\Delta\,\hat{l}_1 \, \hat{l}_2 \,\hat{l}^{\prime}_{h^2}\,
\hat{l}_\Delta
\nonumber \\
& & \nonumber \\
& &(j_{p^1}\, m_{p^1}\, j_{p^2}\,
m_{p^2}\mid J_p \, M_p )\,
(J_p\, M_p\, J^{\prime}_h\, M^{\prime}_h\mid j_t\, m_t)\;
Y_{l_{p^1} m_{l_{p^1}}} (\hat{q}^{\prime}_{p^1} )\,
Y_{l_{p^2} m_{l_{p^2}}} (\hat{q}^{\prime}_{p^2} )
\nonumber\\
& & \nonumber\\
& & (l_{p^1}\, m_{l_{p^1}}\, \ohalf\,
m^{\prime}_{s^1}\mid j_{p^1}\, m_{p^1} )\,
(l_{p^2}\, m_{l_{p^2}}\, \ohalf\,
m^{\prime}_{s^2}\mid j_{p^2}\, m_{p^2} )\;
W(J_2\, j_{p^1}\, j_t\,j_h\, ;\, j_\Delta \, J_1 )
\nonumber\\
& &
(l^{\prime}_{h^2}\, 0 \, l_2 \, 0\mid l_{p^2}\, 0)\;
(l_\Delta \, 0 \, l_1 \, 0\mid l_{p^1}\, 0)\;
\left \{ \begin{array}{ccc} j_{p^1} & j^{\prime}_{h^1} & J_1 \\   \\
                            j_{p^2} & j^{\prime}_{h^2} & J_2 \\   \\
                            J_p     & J^{\prime}_h & j_t
                            \end{array} \right \}
\left \{ \begin{array}{ccc} l_{p^1}  & \ohalf & j_{p^1} \\   \\
                            l_\Delta & \thalf & j_\Delta \\   \\
                            l_1      & 1      & J_2
                            \end{array} \right \}
\left \{ \begin{array}{ccc} l_{p^2}  & \ohalf & j_{p^2} \\   \\
                            l^{\prime}_{h^2} & \ohalf &
                                        j^{\prime}_{h^2} \\   \\
                            l_2      & 1      & J_2
                            \end{array} \right \} \nonumber\\
& & (-\sqrt{3})\,
\sum^{1}_{m=-1}
(1\, m \; \ohalf\, \tau_{p^1}\mid \thalf\, \tau_\Delta )\;
(1\, m \; \ohalf\, \tau^{\prime}_{h^2}\mid \ohalf\, \tau_{p^2} )
\nonumber\\
& & \int dr_2\; r^{2}_{2}\;
\phi_{n^{\prime}_{h^2} \;
(l^{\prime}_{h^2} \, \frac{1}{2}) \, j^{\prime}_{h^2}}(r_2 )\,
\chi_{l_{p^2}}(q^{\prime}_{p^2} r_2 )\;
\int dr_1\; r_1\;
v_{l_1 , l_2 , J_2}(r_1 , r_2)\,
\chi_{l_{p^1}}(q^{\prime}_{p^1} r_1 )\,
\psi^{\; (j_t\, m_t)\,}_{(\Delta\, h)}(r_1 )
\; .
\nonumber\\
\label{BGl3}
\end{eqnarray}
Here $\phi_{n^{\prime}_{h^2} \;
(l^{\prime}_{h^2} \, \frac{1}{2}) \, j^{\prime}_{h^2}}$ denotes the 
radial hole wave function with quantum numbers 
$n^{\prime}_{h^2} \, l^{\prime}_{h^2} \, j^{\prime}_{h^2}$.
The non-local potential $v_{l_1 , l_2 , J_2}(r_1 , r_2)$ 
is the sum of the central and tensor part of the decay 
interaction~\cite{Horie61}:  
\begin{equation}
v_{l_1 , l_2 , J_2}(r_1 , r_2) =
-4\pi\; \delta_{l_1 , J_2}\,\delta_{l_2 , J_2}\;
V^{C}_{J_2}(r_1 , r_2 ) \; +\;
4\pi\, \sqrt{6}\; (-1)^{J_2}\;\hat{l}_{1}\,\hat{l}_{2}\;
(l_1\, 0\, l_2\, 0 \mid 2\, 0)\; W(l_1 \, 1\, l_2\, 1\, ;\, J_2\, 2)\;
V^{T}_{l_1 ,l_2 , J_2}(r_1 , r_2 )\;\; . 
\label{BGl4}
\end{equation}
For the direct matrix element $V^{C}_{J_2}$ and 
$V^{T}_{l_1 ,l_2 , J_2}$ 
are the multipole expanded central and tensor parts of 
$V_{\pi} + V_{\rho} + V_{\delta}$~\cite{Horie61}; for the 
exchange matrix element they are the multipole 
expanded central and tensor parts of $V_{\pi}+V_{\rho}$.


\end{document}